\documentclass[a4paper,fleqn]{cas-dc}
\usepackage[numbers,sort&compress]{natbib}

\ExplSyntaxOn
\keys_define:nn { cas / fig }
  {
    unknown .code:n =
      { \tl_if_blank:nT {#1} { \tl_set:Nx \l_fig_pos_tl { \l_keys_key_tl } } }
  }
\ExplSyntaxOff

\begin{document}
\let\WriteBookmarks\relax
\def\floatpagepagefraction{1}
\def\textpagefraction{.001}

\shorttitle{Intermittency-induced regimes in fast-slow dynamical systems}    

\shortauthors{A. Barone et al.}  

\title [mode = title]{Intermittency-induced regimes in fast-slow dynamical systems}  

%%%%%%%%%%%% AUTORE 1
\author[1]{Alessandro Barone}
\cormark[1]
\ead{alessandro.barone7@unibo.it}
\affiliation[1]{organization={Department of Physics and Astronomy, University of Bologna},
            addressline={Via Carlo Berti Pichat, 6/2}, 
            city={Bologna},
            postcode={40127}, 
            country={Italy}
            }

%%%%%%%%%%%% AUTORE 2
\author[1]{Alberto Carrassi}

%%%%%%%%%%%% AUTORE 3
\author[2]{Jonathan Demaeyer}
\affiliation[2]{organization={Royal Meteorological Institute of Belgium, Meteorological and Climatological Research},
            addressline={Avenue Circulaire, 3}, 
            city={Brussels},
            postcode={1180}, 
            country={Belgium}
            }

%%%%%%%%%%%% AUTORE 4
\author[2,3]{St\'ephane Vannitsem}
\affiliation[3]{organization={School of Physical \& Mathematical Sciences, Nanyang Technological University},
            country={Singapore}
            }

\cortext[1]{Corresponding author}

\begin{abstract}
Intermittency is ubiquitous in nature and often arises from interactions between processes evolving on different time scales. We investigate whether, and how, intermittency in a fast system can be transmitted to a slower system that, in isolation, would settle onto a periodic regime. We address this in a unidirectionally coupled (skew-product) setting using two models of increasing complexity: coupled Lorenz-63 systems and coupled Kuramoto–Sivashinsky equations, in which the fast driving subsystem exhibits type-I intermittency. Increasing the driver’s burst frequency progressively displaces the response attractor from its unforced limit cycle, until saturation. The induced regimes are jointly controlled by the coupling strength $C$ and the time-scale separation $\tau$, which we explore through the variance of ensemble maxima, power spectra, extreme value statistics and phase-locking indices. In the Lorenz-63 model, $C$ mainly governs the loss of periodicity and the statistics of extremes, while $\tau$ mainly governs the reshaping of the response spectrum; in the Kuramoto–Sivashinsky model these roles are less separable. In both systems, the transition to the intermittency-dominated regime is marked by heavy-tailed (Fréchet-type) extremes, which give way to bounded statistics at stronger coupling. The response delay, estimated via time-lagged mutual information, decreases exponentially as the two time scales become comparable. We identify phase synchronization during the driver’s laminar phases as the mechanism behind several regime transitions, with chaotic bursts repeatedly breaking an otherwise stable locking. These results shed light on how fast intermittent variability, such as atmospheric forcing, may shape slower climate components such as the ocean or sea ice.
\end{abstract}

\begin{highlights}
\item A fast intermittent driver induces intermittency in a slow periodic response system
\item Heavy-tailed extremes mark the transition from laminar to intermittent regimes
\item Several regime transitions are explainable through a phase-locking mechanism 
\item Response delay scales exponentially with the fast–slow time-scale separation
\end{highlights}

% Keywords

\maketitle

%% ====================================================================
\section{Introduction}\label{Intro}

Among the many peculiar properties of nonlinear dynamical systems, intermittency is one of the most widespread. It appears in systems of very different nature and complexity. Both historical and recent studies have highlighted the central role of intermittent signals in natural phenomena; examples can be found in medicine, neuroscience, economics, and ecology (see \cite{elaskar_new_2017} and references therein). In the field of Earth and geophysical sciences, intermittency is encountered in several different guises. In some cases, it is understood in the dynamical-systems (DS) sense \citep{pomeau_intermittent_1980, manneville_intermittency_1979} adopted throughout this work, or identified through experimental observations of spatio-temporal intermittency in convective flows \citep{berge_intermittency_1980, daviaud_spatio-temporal_1989}, as well as across multifractal and scaling frameworks used to characterize the scale-invariant structure of geophysical fields such as rainfall and sea-ice deformation \citep{lovejoy_weather_2013, schmitt_modeling_1998, rampal_multi-fractal_2019, pandey_high_2026}. More broadly, the term is also used to denote the whole class of meteorological and climatological phenomena that exhibit an apparently random alternation between distinct states, such as intermittent oceanic mixing \cite{manucharyan_climate_2011} or the intermittent correlations in some climate modes \citep{bellucci_intermittent_2022}. Recently, intermittency has also been investigated from a predictive perspective, focusing on the identification of real-time indicators and precursors capable of anticipating abrupt in-time transitions in dynamical systems for fixed parameter values or external forcings \cite{barone_structural_2025}.

The Earth system is characterized by the interplay of different temporal and spatial scales \citep{ghil_lucarini_physics_2020}. Climate variability is a prime example of the importance of such multiscale interactions, with its spatial and temporal structure changing markedly across scales. This is reflected both in the fingerprint left on the variance spectrum \citep{mitchell_overview_1976, ghil_natural_2002, von_der_heydt_quantification_2020} as well as in the power-law behavior of its scaling properties \citep{franzke_structure_2020}. In such multiscale systems, the interaction between fast and slow components is often crucial. In particular, the fast subsystem may exhibit intermittent dynamics, while the slower subsystem may evolve on quasi-stationary or weakly oscillatory regimes. A concrete illustration of this kind of interaction is offered by the atmosphere--ocean system, where fast, intermittent atmospheric forcing can leave a durable imprint on the slower oceanic component. In the stochastic climate framework introduced in the seminal work of Hasselmann \cite{Hasselmann1976} and subsequently applied to sea-surface temperature anomalies \cite{Frankignoul1977}, the fast atmospheric variability is treated as a stochastic forcing acting on the slow ocean, which integrates it in time and responds with red-noise fluctuations on much longer time scales. When the fast component is intermittent, however, the effective forcing on the slow one is no longer Gaussian: deterministic homogenisation of fast--slow systems shows that intermittent fast dynamics induce non-Gaussian $\alpha$-stable (L\'evy) forcing on the slow variables \cite{gottwald_model_2021}. This mechanism has been invoked to explain abrupt glacial climate changes \cite{gottwald_model_2021,riechers_glacial_2024} and to represent westerly wind bursts in conceptual models of ENSO \cite{gottwald_improved_2026}.
In tropical cyclones, despite the short duration of each individual event, cyclone-induced intermittent mixing has been shown to produce robust, long-lasting changes in sea surface and subsurface temperature, ocean circulation, and poleward heat transport \citep{manucharyan_climate_2011}. While such studies have focused on the broader effect that an intermittent forcing, often represented stochastically, can produce on the forced system, here we ask two distinct questions: can intermittent behavior in a fast component induce intermittency in a slower one? And if so, how is this information transferred across scales?

We answer these questions by investigating the dynamics of coupled fast--slow systems in which a fast intermittent system forces a slower subsystem that, in isolation, would exhibit a regular behavior, namely a limit cycle. We use an idealized configuration in which there is no feedback from the slow to the fast component. In this setting, the fast system acts as an external intermittent forcing on the slow one. Conceptually, this framework can be interpreted as a simplified representation of a slow natural cycle, such as an oceanic mode of variability, being forced by a fast intermittent atmospheric component.

The systems we consider are two idealized models of increasing complexity: two coupled Lorenz-63 (L63) systems \cite{lorenz_deterministic_1963}, following the formulation proposed in \cite{kalnay_breeding_2003}, in which the feedback term is suppressed; and a coupled spatially extended system based on the Kuramoto--Sivashinsky equation (KSE) \cite{kuramoto_persistent_1976,sivashinsky_nonlinear_1977}, following the coupled structure introduced in \cite{evensen_iterative_2024}. The Kuramoto--Sivashinsky equation is a paradigm of a spatially-extended turbulent system that has been extensively used as an idealized testbed for exploring the nature of turbulence and for testing techniques that could be difficult to implement in more realistic systems \cite[e.g.][]{Manneville1981,Nicolaenko1985,Hyman1986,Manneville1990, Vannitsem1994,Vannitsem1995,Jardak2010,evensen_iterative_2024}.

The overarching objective of this work is to understand how intermittent information can be transmitted between systems evolving on different time scales, and under which conditions intermittency can be induced in the slower subsystem. Beyond its general relevance for multiscale nonlinear dynamics, this study is motivated by a conceptual climate interpretation: an intermittent atmospheric component forcing a slower oceanic system in near-stationary conditions. 

The paper is structured as follows. In Section~\ref{sec:setup} we describe the general unidirectional fast--slow forcing framework. In Section~\ref{sec:transitions} we investigate the transitions induced in the response system by the intermittent driver. In Section~\ref{SEC:RESP} we examine the response delay and its scaling behavior. In Section~\ref{sec:phase} we analyze the phase locking between the two subsystems and its role in shaping part of the transitions observed. Finally, in Section~\ref{sec:concl}, we summarize the main findings and discuss possible future directions of this work. The settings of all the numerical experiments are collected in Table~\ref{tab:experiments}.

%% ====================================================================
\section{Prototypical uni-directionally forced fast-slow dynamics} \label{sec:setup}

We consider a general class of systems consisting of two interacting subsystems, with a fast subsystem forcing a slower one:

\begin{equation}
\label{eq:full_system}
\dot{\mathbf{X}} =
\begin{pmatrix}
\mathbf{F}(\mathbf{X}_1, p) \\
\tau \left[ \mathbf{G}(\mathbf{X}_2) + \mathcal{C}(\mathbf{X}_1, \mathbf{X}_2) \right]
\end{pmatrix},
\qquad
\mathbf{X} =
\begin{pmatrix}
\mathbf{X}_1 \\
\mathbf{X}_2
\end{pmatrix}.
\end{equation}

Equation~\eqref{eq:full_system} is written in a general form to encompass both systems considered in this work, where the coupling term, $\mathcal{C}(\mathbf{X}_1, \mathbf{X}_2)$, is linear and additive. For the coupled L63 model, $\mathbf{F}$ and $\mathbf{G}$ are ordinary differential equations (ODEs) and $\mathbf{X}_1, \mathbf{X}_2  \in \mathbb{R}^N$ are the state vectors. For the coupled KSE, which is described by partial differential equations (PDEs), $\mathbf{X}_1, \mathbf{X}_2  \in \mathbb{R}^N$ are instead to be understood as the state vectors obtained after the spatial discretization. Here $\tau>0$ is the ratio between the characteristic time scales of the two subsystems: smaller values of $\tau$ correspond to a larger time-scale separation, $\tau=1$ to no separation, and $\tau>1$ to a response faster than the driver. $p\in\mathbb{R}$ is the control parameter of the forcing subsystem.

Since $\mathbf{F}$ does not depend on $\mathbf{X}_2$, Eq.~\eqref{eq:full_system} is a skew-product system: $\mathbf{X}_1$ evolves autonomously and acts as an external, possibly intermittent, driving force on $\mathbf{X}_2$, without any feedback from the response.

\subsection{Coupled Lorenz 63}

We consider two Lorenz-63 systems coupled as shown in Eq.~\eqref{eq:l63}:

\begin{equation}
\label{eq:l63}
\left\{
    \begin{aligned}
    \dot{x}_1 \;&=\; \sigma (y_1 - x_1) \\
    \dot{y}_1 \;&=\; x_1 (\rho_1 - z_1) - y_1 \\
    \dot{z}_1 \;&=\; x_1 y_1 - \beta z_1 \\[0.8em]
    \dot{x}_2 \;&=\; \tau \Bigl[ \sigma (y_2 - x_2) - C x_1 \Bigr] \\
    \dot{y}_2 \;&=\; \tau \Bigl[ x_2 (\rho_2 - z_2) - y_2 + C y_1 \Bigr] \\
    \dot{z}_2 \;&=\; \tau \Bigl[ x_2 y_2 - \beta z_2 + C z_1 \Bigr] .
    \end{aligned}
\right.
\end{equation}

Our unidirectional coupled configuration in Eq.~\eqref{eq:l63} is a special case of the two-way coupled version introduced in \cite{kalnay_breeding_2003} and later employed as an idealized testbed for coupled climate prediction and data assimilation studies \cite{carrassi2014,weber2015,garciaoliva2025}.

Both subsystems share the classical L63 parameters $\sigma=10$ and $\beta=8/3$, while the Rayleigh parameter is fixed at $\rho_1 = 166.2$ for the fast subsystem and $\rho_2 = 166.0$ for the slow one. This choice places the fast subsystem in a type-I intermittent regime \cite{pomeau_intermittent_1980,manneville_intermittency_1979}, in which trajectories alternate between chaotic bursts around the two symmetric fixed points and long, nearly periodic laminar phases. For $\rho_2 = 166.0$, in the absence of coupling to the fast subsystem, the slow subsystem would instead settle onto a stable period-2 limit cycle. We shall see that it is driven into an intermittent regime once forced by $\mathbf{X}_1$ through the coupling term $\mathcal{C}$.

The coupled system, Eq.~\eqref{eq:l63}, is integrated with a fixed-step, fourth-order Runge--Kutta scheme (RK4), with time step $\Delta t = 0.01$. % CHECK: dt = 0.01 inferred from the variance-of-maxima run (100000 steps over 1000 TU); confirm it is the same for all L63 experiments

\subsection{Coupled Kuramoto--Sivashinsky equations}

The coupled version of the KSE used here has been introduced by \cite{evensen_iterative_2024} and reads: 

\begin{equation}
\label{eq:kse}
\left \{
    \begin{aligned}
    \frac{\partial u_1}{\partial t} \;&=\; \frac{\partial^2 u_1}{\partial x^2} - \nu_1 \frac{\partial^4 u_1}{\partial x^4} - u_1 \frac{\partial u_1}{\partial x}\\
    \frac{\partial u_2}{\partial t} \;&=\; \tau \left[ \frac{\partial^2 u_2}{\partial x^2} - \nu_2 \frac{\partial^4 u_2}{\partial x^4} - u_2 \frac{\partial u_2}{\partial x} + C\cdot (u_1 - u_2) \right],
    \end{aligned}
\right.
\end{equation}
with $u_1(x,t)$ and $u_2(x,t)$ representing, respectively, the solution of the forcing and the response system.

\begin{table*}[t]
\centering
\caption{Sequence of the numerical experiments. Unless otherwise stated, $\rho_1=166.2$, $\rho_2=166$ (L63) and $\nu_1=0.92$, $\nu_2=1$ (KSE). Time steps: $\Delta t=0.01$ (L63, RK4) and $\Delta t=0.1$ (KSE, ETDRK4); KSE domain length $L=10 \times \pi$, $N=101$ grid points. $T$: total integration time; $T_{\mathrm{tr}}$: discarded transient, both in TU.}
\label{tab:experiments}
\footnotesize
\setlength{\tabcolsep}{3pt}
\newcolumntype{P}[1]{>{\raggedright\arraybackslash}p{#1}}
\begin{tabular}{@{}P{1.6cm} P{4.0cm} P{1.9cm} P{2.6cm} P{6.4cm}@{}}
\toprule
Model & Parameters & $T$ / $T_{\mathrm{tr}}$ & Ensemble & Method settings \\
\midrule
\multicolumn{5}{@{}c@{}}{\textbf{Sliced Wasserstein distance}} \\
L63 & $\rho_1\in[166,168]$, 20 values; $C=0.5$, $\tau=0.1$ & $10^5$ / $10^4$ & 1 IC, $\mathcal{U}(-10,10)$ & subsampling every 10 steps; reference $\rho_1=\rho_2$; $n_{\mathrm{proj}}=100$ \\
KSE & $\nu_1\in[0.86,1]$, 20 values; $C=0.2$, $\tau=0.1$ & $10^4$ / $10^3$ & 1 IC (fixed) & reference $\nu_1=\nu_2$; $n_{\mathrm{proj}}=100$ \\
\midrule
\multicolumn{5}{@{}c@{}}{\textbf{Variance of maxima}} \\
L63 & $\tau, C\in[0.1,1.1)$ 
 & $2\times10^4$ / $1.9\times10^4$ & 20 ICs, $\mathcal{U}(-5,5)$ & maxima sampled in the last $500$ TU; variance of the maxima across the ensemble \\
KSE & $\tau, C\in[0.01,1]$
 & $1.1\times10^4$ / $10^4$ & 20 ICs, $u_1=0.1\,\mathcal{U}(0,1)$, $u_2=u_1+10^{-3}$ 
 & maxima sampled in the last $500$ TU; variance of the maxima across the ensemble \\
\midrule
\multicolumn{5}{@{}c@{}}{\textbf{Power spectra}} \\
L63 & $\tau, C \in [0.01,1]$ 
 & $5\times10^3$ / $10^3$
 & 1 IC & RMSE between $\log_{10}$ PSDs of $\mathbf{X}_1$ and $\mathbf{X}_2$ \\
KSE & $\tau, C \in [0.01,1]$ 
 & $5\times10^3$ / $10^3$
 & 1 IC & RMSE between $\log_{10}$ PSDs of $\mathbf{X}_1$ and $\mathbf{X}_2$ \\
\midrule
\multicolumn{5}{@{}c@{}}{\textbf{Decorrelation time and extreme values}} \\
L63, $\tau_d$ & $\tau, C\in[0.01,1]$, & $10^4$ / $5 \times 10^3$ & 1 IC & ACF below $1/e$ for at least $400$ TU \\
L63, GEV & as above & $10^5$ / $5 \times 10^3$ & 1 IC & $\Delta T=2\tau_d$, $\Delta T\ge20$; at least 30 maxima; MLE fit \\
KSE, $\tau_d$ & $\tau, C\in[0.01,1]$, & $2\times10^3$ / $2\times10^3$ & 1 IC & ACF below $1/e$ for at least $40$ TU \\
KSE, GEV & as above & $2\times10^5$ / $2\times10^3$ & 1 IC & $\Delta T=2\tau_d$, $20\le\Delta T\le2000$; at least 20 maxima; MLE fit \\
\midrule
\multicolumn{5}{@{}c@{}}{\textbf{Response delay}} \\
L63, $R(t)$ & $C=1$, $\tau=0.59$ & $5 \times 10^3$ / $10^3$ 
 & 1 IC & Hilbert phase; sliding window $w=5$ TU \\
KSE, $R(t)$ & $C=0.2$, $\tau=0.05$ & $10^4$ / $2\times10^3$ % TODO: inferred from Fig. 5 axis
 & 1 IC & as for L63 \\
L63, MI & $\tau \in [0.01, 1]$; $C=0.5$ 
 & % TODO
 & 50 ICs % TODO
 & min--max normalised signals; $l^*$ maximising $I(l)$ after Gaussian smoothing; weighted exponential fit of $\log(-l^*)$ \\
KSE, MI & $\tau\in[0.015,0.5]$; $C=0.5$ & $6\times10^4$ / $4 \times 10^3$ & 50 ICs, $u_1+0.01\,\mathcal{N}(0,1)$ & as for L63, 50 bins, $|l|\le150$ TU
 \\
\midrule
\multicolumn{5}{@{}c@{}}{\textbf{Phase locking}} \\
L63 & $\tau, C \in [0.01,1.5]$; $\rho_1=166.2$ and periodic driver $\rho_1=166$ 
 & $10^4$ / $4 \times 10^3$ 
 & 1 IC & time-averaged $\overline{R}$ over the whole trajectory \\
KSE & $\tau=C/\tan\theta$, $\theta\in\{22^\circ,45^\circ,67^\circ\}$ & 
 & 1 IC & time-averaged $\overline{R}$ along each direction \\
\bottomrule
\end{tabular}
\end{table*}

Following \cite{barone_structural_2025}, we select values of the fast-system viscosity parameter, $\nu_1$, such that it behaves intermittently. In particular, the dynamics is characterized by alternations between chaotic bursts and nearly periodic traveling-wave states.

% CHECK: text says u_1 ~ U(0,1), table (variance of maxima) says 0.1 U(0,1); align with the code
The fast field is initialized as $u_1(x,t=0) \sim \mathcal{U}(0,1)$, sampled once and for all, while the slow field is initialized as a slightly perturbed copy of it, $u_2(x,0) = u_1(x,0) + \delta$, with $\delta = 10^{-3}$. This initialization strategy is motivated by the results in \cite{barone_symmetry-driven_2026}, where it was shown that the KSE, for identical parameter values, displays the coexistence of multiple attractors: depending on the initial condition, the system converges to distinct (strange) attractors or periodic traveling-wave solutions. By fixing the same initial condition $u_1$ across all simulations, we ensure that the fast subsystem always evolves toward the same invariant intermittent attractor. This in turn guarantees that the slow subsystem is consistently forced by the same underlying intermittent dynamics, characterized by alternating chaotic phases and coherent traveling-wave states in physical space. This strategy is adopted whenever a single trajectory is integrated. When an ensemble of trajectories is required (see Table~\ref{tab:experiments}), the members are instead generated by adding a small random perturbation to the fixed initial condition $u_1(x,0)$, in order to keep the trajectories as close as possible to the same intermittent attractor. Nevertheless, owing to the coexistence of multiple attractors discussed in \cite{barone_symmetry-driven_2026}, ensemble members may converge to distinct invariant sets, and the resulting statistics are inevitably affected by this multistable structure of the KSE.

The dynamics is written in Fourier space as a system of coupled ODEs for the Fourier coefficients, on a periodic domain of length $L= 10 \times \pi$ discretized with $N= 101$ grid points. The integration is carried out using the fourth-order exponential time-differencing Runge--Kutta scheme (ETDRK4) with time step $\Delta t = 0.1$, following the formulation introduced by Cox and Matthews \cite{cox_exponential_2002} and further refined by Kassam and Trefethen \cite{kassam_fourth-order_2005}. This approach ensures numerical stability and accuracy for stiff dissipative systems such as the KSE.

%% ====================================================================
\section{Inducing intermittency and transitions} \label{sec:transitions}

As a first experiment, we investigate how the frequency of the chaotic bursts occurring in the fast subsystem $\mathbf{X}_1$ progressively induces modifications in the variability of $\mathbf{X}_2$ away from its own natural limit-cycle (or traveling-wave, in the KSE case) attractor. To quantify this effect, we compute the sliced Wasserstein distance (SWD) $\mathcal{D}$ between the attractor induced in $\mathbf{X}_2$ by the coupling, and the one the slow system would have in the absence of the coupling, $\mathbf{X}_{ref}$ (see Appendix~\ref{app:sliced-wd} for details).

For both systems, the coupling strength $C$ and the time-scale separation $\tau$ are held fixed, while the parameters that structurally control the dynamical regime of the driver, $\rho_1$ and $\nu_1$ respectively for the coupled L63 and KSE, are varied.
Studying the deviation from the uncoupled slow system attractor against the control parameters is motivated by the fact that both the L63 and KSE fast subsystems, in the parameter regions considered, exhibit Type-I intermittency \cite{pomeau_intermittent_1980,manneville_intermittency_1979,barone_structural_2025}. Near the bifurcation point the statistics of the laminar phases in Type-I intermittency follow two well-defined, reciprocal power laws: the average laminar length $\langle \ell \rangle$ diverges as the control parameter approaches its critical value, while the burst frequency vanishes correspondingly \cite{elaskar_new_2017,manneville_intermittency_1979,barone_structural_2025}. As a consequence, there exists a well-defined, monotonic relationship between $\rho_1$ (resp.\ $\nu_1$) and the average length of the laminar phase, which justifies using $\rho_1$ (resp.\ $\nu_1$) directly as a proxy for burst frequency.

\begin{figure*}[t]
\centering
\includegraphics[width=0.9\linewidth]{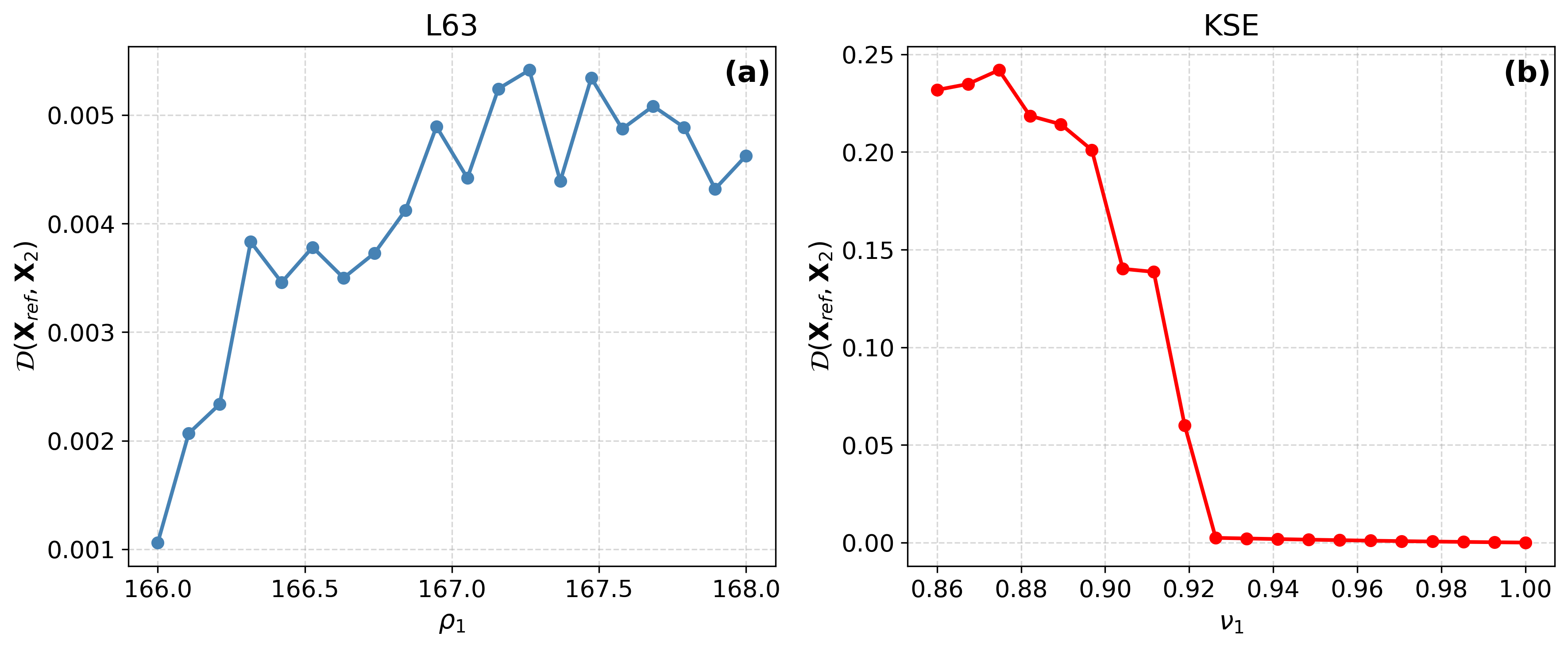}
\caption{Sliced Wasserstein distance $\mathcal{D}(\mathbf{X}_{ref},\mathbf{X}_2)$ between the attractor induced in the response subsystem and its unforced reference attractor, as a function of the parameter controlling the driver's burst frequency. (a) L63 system, as a function of the fast subsystem's Rayleigh parameter $\rho_1$; the reference is the unforced limit cycle ($\rho_1=\rho_2$), and the onset of intermittency in the driver is at $\rho_1\approx166.07$. (b) KSE, as a function of the fast-subsystem viscosity $\nu_1$ (to be read from right to left, since decreasing $\nu_1$ corresponds to increasing burst frequency); the reference is the unforced travelling-wave regime ($\nu_1=\nu_2$). Parameters are given in Table~\ref{tab:experiments}.}
\label{fig:WD}
\end{figure*}

\paragraph{{\bf L63} -} The coupling strength, the time-scale separation and the slow subsystem's bifurcation parameter $\rho_2$ are held fixed, while the bifurcation parameter of the fast subsystem, $\rho_1$, is varied across a range encompassing the transition from the periodic regime through the onset of intermittency at $\rho_1 \approx 166.07$ into increasingly chaotic driving. For each value of $\rho_1$, a long trajectory is integrated and, after discarding an initial transient, subsampled to build the dataset. The dataset is made up directly from the response subsystem's state vector $\mathbf{X}_2 = (x_2, y_2, z_2)$, pooled over the sampled trajectory and standardized to zero mean and unit variance component-wise. The reference (unforced) case corresponds to $\rho_1 = \rho_2$.

\paragraph{{\bf KSE} -} The slow-system viscosity $\nu_2$, the coupling strength and the time-scale separation are held fixed, while the fast-system viscosity $\nu_1$ is varied. Each simulation starts from the same fixed initial condition, and an initial transient is discarded; we verified that it is sufficient in all cases by tracking $\|\mathbf{X}_2(t+\Delta t)-\mathbf{X}_2(t)\|_2$ for all values of $\nu_1$. The reference (unforced) case corresponds to $\nu_1 = \nu_2$, for which $\mathbf{X}_2$ evolves in its natural traveling-wave (limit-cycle) regime, undisturbed by any effective intermittent forcing. For each $\nu_1$, the response trajectory $u_2(x,t)$ is projected onto the local phase space of its first and second spatial derivatives $(u_x, u_{xx})$. The resulting fields are evaluated at every grid point and post-transient time step, pooled together and standardized, yielding an empirical distribution in the $(u_x, u_{xx})$ plane. The sliced Wasserstein distance (Appendix~\ref{app:sliced-wd}) between this point cloud and the reference gives the curve $\mathcal{D}(\nu_1)$ in Fig.~\ref{fig:WD}b. For details on the numerical experiments see Table~\ref{tab:experiments}.

\paragraph{{\bf Results} -} From Fig.~\ref{fig:WD}a we see that, in the L63 system, $\mathcal{D}(\mathbf{X}_{ref}, \mathbf{X}_2)$ increases with $\rho_1$. This is because the frequency of chaotic bursts in the driving subsystem $\mathbf{X}_1$ grows with $\rho_1$, which reflects a structural change induced in the attractor of the slow system: the more frequently $\mathbf{X}_1$ bursts into its chaotic phase, the further the response subsystem $\mathbf{X}_2$ is pushed away from its uncoupled configuration. Beyond $\rho_1 \approx 167$, however, the distance ceases to grow monotonically and saturates, oscillating around a roughly constant mean value. This suggests that, for the configuration of $C$ and $\tau$ used here, a further increase in the burst frequency of the driver no longer produces further modification of the driven system, {\it i.e.} its attractor has effectively reached its maximal configuration. We interpret this saturation as a signature of the slow system's own memory and relaxation capacity, set by $\tau$, which bounds how far the induced distribution can be deformed regardless of how frequently the driver bursts.

In the KSE (Fig.~\ref{fig:WD}b) the distance $\mathcal{D}(\mathbf{X}_{ref}, \mathbf{X}_2)$ is plotted against the fast-subsystem viscosity $\nu_1$; the plot is to be read from right to left, since decreasing $\nu_1$ corresponds to increasing burst frequency in the driver \cite{barone_structural_2025,barone_symmetry-driven_2026}. Starting from the onset of intermittency ($\nu_1 \approx 0.92$--$0.93$), $\mathcal{D}$ rises sharply as $\nu_1$ decreases: the growing frequency of chaotic bursts induces a structural change in the regime of $\mathbf{X}_2$, measurable as the distance between the two distributions. This growth is bounded: for sufficiently low values of $\nu_1$ the distance saturates. Both systems therefore display a progressive, bounded displacement of the response attractor with the burst frequency of the driver.

%% --------------------------------------------------------------------
\subsection{The interplay of the time scale difference and the coupling strength} \label{sec:tauc}

Given the structure of the systems, the emergence of new types of solutions depends crucially on the parameters $\tau$ and $C$. We expect them to have an impact on the intermittent behavior of the coupled system, and even on its possible occurrence. Here, $\tau$ represents the time-scale separation between the two subsystems, therefore controlling how the slow component responds to intermittent bursts in the fast dynamics, while $C$ directly determines the strength of the forcing exerted on the slow subsystem. We use two diagnostics, both computed from a scalar observable of the response subsystem: the variance of its sampled local maxima, an amplitude-based statistic, and its power spectral density (PSD), which characterizes how the energy is distributed across frequencies. A third diagnostic, based on extreme value theory, is presented in Sect.~\ref{sec:ext}. For details on the numerical experiments see Table~\ref{tab:experiments}.

\paragraph{{\bf L63} -} The observable is $x_2(t)$. In analogy with a bifurcation diagram (also referred to as an orbit diagram), for each $(\tau,C)$ we initialize an ensemble of random initial conditions and, for each realization, we sample the local maxima of $x_2$ over a time window taken at the end of a long simulation. In the standard case of a single-parameter variation, one would directly plot these maxima along the vertical axis of the bifurcation diagram. Here, instead, we consider the variance of the sampled maxima across the ensemble as a proxy to quantify their dispersion. This allows us to identify regions where the maxima are either widely spread or nearly coincident. The underlying hypothesis is that regions of the $(\tau,C)$ plane characterized by large variance correspond to significant changes in the system's dynamics, including regime transitions or structural modifications of the attractor. The spectral reshaping is quantified by the RMSE between the $\log_{10}$ PSDs of the driver and of the response, as a function of $\tau$ and $C$, and by the response PSD at fixed $\tau$ for varying $C$.

\paragraph{{\bf KSE} -} The same approach is applied to the KSE, with some adaptations dictated by its higher computational cost and spatially extended nature. The observable is the instantaneous energy of the response field, $\varepsilon_2(t) = \lVert u_2(x,t)\rVert_{L^2}$, the local maxima of which are sampled over a window at the end of a long simulation, for an ensemble of randomly perturbed initial conditions of the driver and response fields. As for the L63, the variance of the sampled maxima across the ensemble is used as a proxy to detect regime transitions and structural changes in the induced attractor, and the PSD analysis is performed on $\varepsilon_2(t)$ and on the corresponding driver observable.

\begin{figure*}[t]
    \centering
    \includegraphics[width=0.9\textwidth]{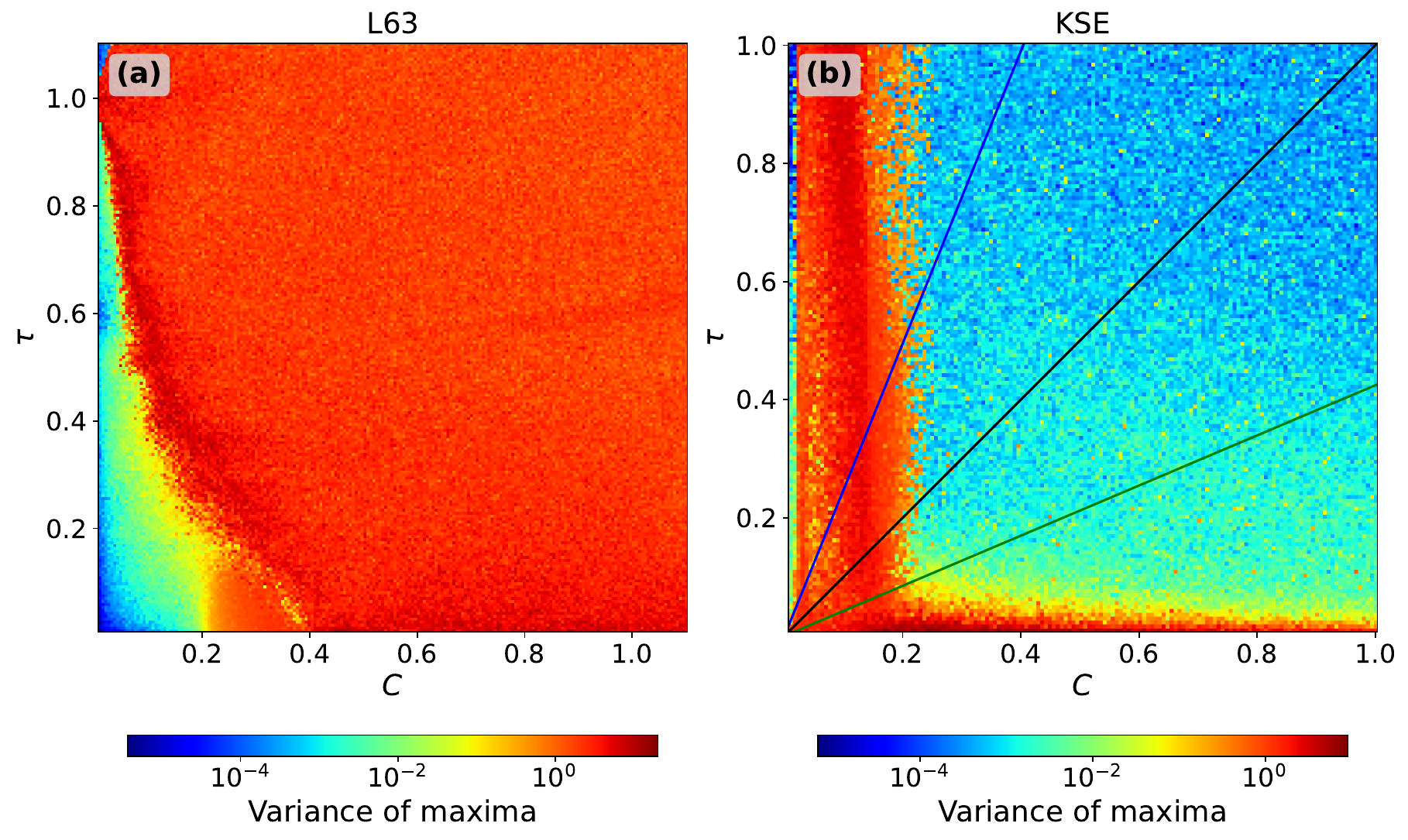}
    \caption{Variance of the sampled local maxima of the response subsystem, used as a proxy for regime transitions and structural changes in the induced attractor, over the $(\tau, C)$ parameter plane, for (a) L63 (maxima of $x_2(t)$) and (b) KSE (maxima of the response energy $\varepsilon_2(t) = \lVert u_2(x,t)\rVert_{L^2}$). Diagonal lines in (b) mark the three directions $\tau = C/\tan\theta$ ($\theta=22^\circ,45^\circ,67^\circ$) used in the phase-locking analysis of Sect.~\ref{sec:phase}. Details are given in Table~\ref{tab:experiments}.}
    \label{fig:C2tau1}
\end{figure*}

\begin{figure*}[t]
    \centering
    \includegraphics[width=\linewidth]{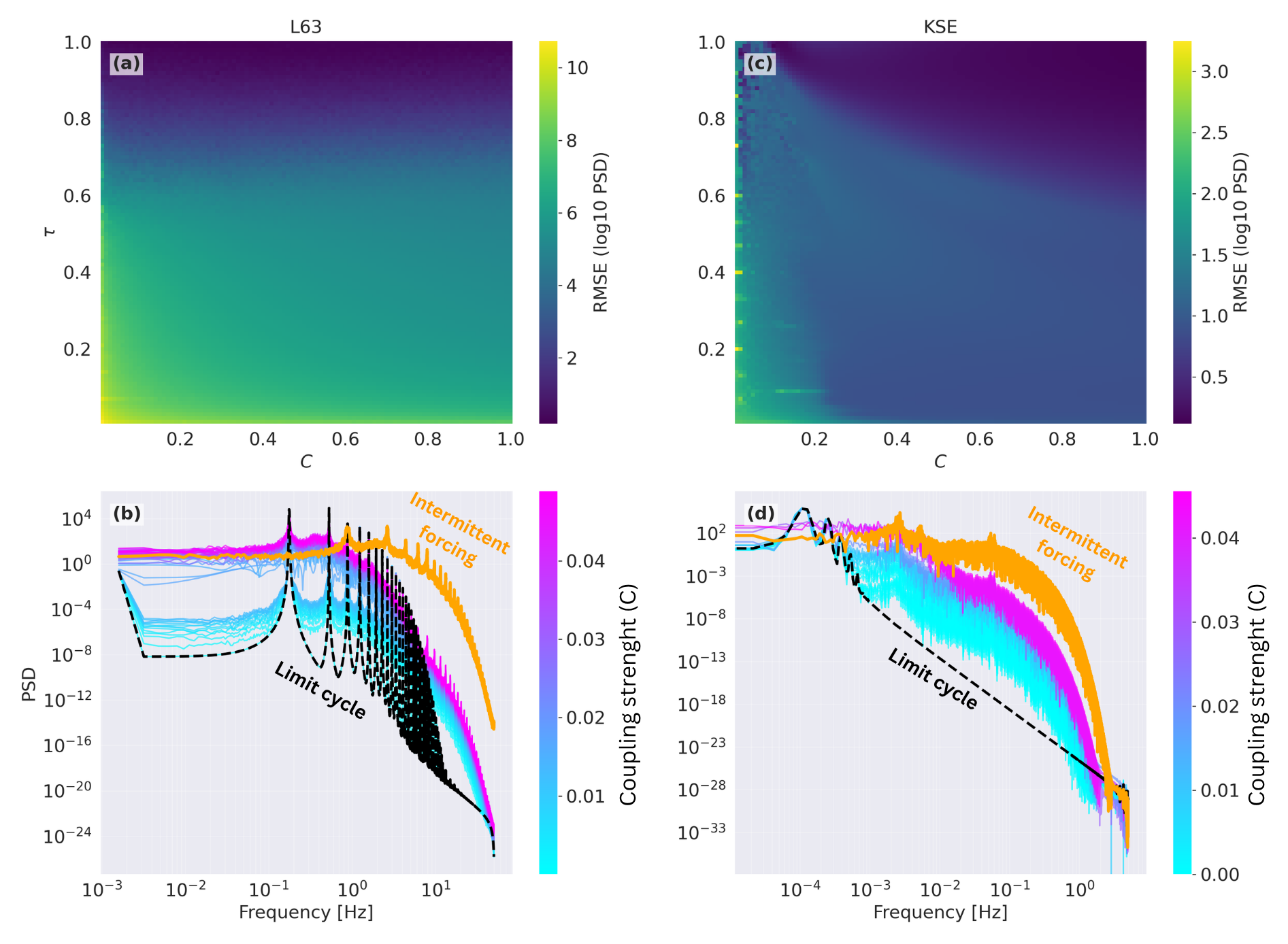}
    \caption{Spectral reshaping of the response subsystem under intermittent forcing. (a) RMSE between the power spectral density (PSD) of the driver $\mathbf{X}_1$ and of the response $\mathbf{X}_2$, as a function of $\tau$ and $C$, for the L63 system. (b) PSD of the L63 response $x_2(t)$ for varying coupling strength $C$ (color scale) at fixed $\tau = 0.2$, together with the PSD of the intermittent driver (orange) and of the unforced limit cycle (dashed black). (c)--(d) Same as (a)--(b) for the KSE system, with the response PSD computed from the energy $\varepsilon_2(t) = \lVert u_2(x,t)\rVert_{L^2}$ and $\tau = 0.05$.}
    \label{fig:spec}
\end{figure*}

\paragraph{{\bf Results} -} In the L63 case (Fig.~\ref{fig:C2tau1}a) the transition to chaos occurs rather progressively as both $\tau$ and $C$ increase. Interestingly, even for very small values of $C$, large variability ({\it i.e.}, proxies for chaos) can be induced by solely reducing the time-scale separation between the two subsystems (recall that larger $\tau$ means shorter time-scale separation, with $\tau=1$ implying no time-scale separation). This suggests the following physical interpretation: the time-scale separation buffers and mitigates abrupt perturbations originating in the fast dynamics, with the effects weakening as the separation is reduced. Moreover, the transition driven by varying $C$ is sharper than that observed when varying $\tau$, indicating that the coupling strength is a more effective tunable parameter to achieve a given transition in the driven slow system.

Figure~\ref{fig:C2tau1}a reveals other interesting features: two faint stripe-like structures of enhanced variance, the first visible just below $\tau \approx 0.6$ and in the range $C \in [0.6, 1.1]$. Although weak, these bands may indicate regime transitions in the coupled dynamics. The second emerges when the time-scale separation is reversed, {\it i.e.}, when the slow subsystem becomes faster than the fast one, and develops with approximately the same slope as the first. A closer inspection of this behavior is reported in the {\it Supplementary Material}, where two additional experiments are performed: one in which the driver is replaced by a purely chaotic dynamics, and another in which it is set to a fully periodic regime. Strikingly, the stripe-like structures become markedly more pronounced in the periodic case, whereas they are absent under chaotic forcing. We therefore conjecture that their appearance is a manifestation of the two subsystems being synchronized; we shall discuss this in further detail in Sect.~\ref{sec:phase}.

The spectral analysis gives a complementary view of the changes induced in the response system. Figure~\ref{fig:spec}a shows the RMSE between the PSD of the intermittent driver $\mathbf{X}_1$ and that of the response $\mathbf{X}_2$. Both parameters contribute to shaping the response spectrum toward that of the intermittent driver, but the dominant role is played by $\tau$, as indicated by the strong gradient along the vertical axis. In Fig.~\ref{fig:spec}b we examine the changes of the spectrum by fixing $\tau$ and varying $C$ alone. For weak coupling, the response spectrum deviates from that of the unperturbed periodic orbit (dashed black curve) while retaining the same principal oscillation frequencies. As the coupling strength increases, the spectrum progressively takes the shape of the driver's one (orange curve), pointing to a gradual transfer of spectral properties from the fast, intermittent subsystem to the slow one.

For the KSE (Fig.~\ref{fig:C2tau1}b) the picture is markedly different. For $C$ close to zero, a narrow low-variance region is observed, where the response essentially retains its unforced behavior. As the coupling increases, a broad high-variance region develops for $C \lesssim 0.25$ over the whole range of $\tau$: here the forcing destabilizes the KSE dynamics, inducing large fluctuations in the amplitude of the maxima (see also the {\it Supplementary Material}). For larger $C$ the variance drops by several orders of magnitude, except for a strip at small $\tau$ that remains at high variance and persists across nearly the entire range of $C$. In the spectral analysis (Fig.~\ref{fig:spec}c), the interplay between $\tau$ and $C$ is markedly less trivial: the coupling strength plays a more prominent role, and the $(\tau, C)$ plane exhibits distinct transition bands with different slopes, rather than a single dominant gradient along one axis. Panel~(d) displays the spectral shaping obtained by varying $C$ alone at fixed $\tau$: the transition occurs in a smooth and regular fashion, and for sufficiently small $C$ the response spectrum already appears almost fully reshaped toward that of the intermittent driver. At low coupling both subsystems initially display peaks at the same oscillation modes as the underlying limit cycle, even as new frequencies are introduced; these peaks are then progressively abandoned as $C$ increases, with the spectrum settling onto the frequencies characteristic of the intermittent dynamics.

The two diagnostics probe different aspects of the response. The variance of the maxima is an amplitude-based statistic and, in the L63, is mainly controlled by $C$, which sets the intensity of the forcing. The PSD instead characterizes how the energy is distributed across frequencies, and in the L63 it is more sensitive to $\tau$, which directly sets the frequencies at which the driver acts on $\mathbf{X}_2$ relative to its intrinsic frequencies. The two diagnostics are therefore complementary: $C$ mainly determines how strongly the response is perturbed, while $\tau$ mainly determines how its temporal structure is reshaped. In the KSE this separation of roles is less clean, with both parameters contributing to both diagnostics.

%% --------------------------------------------------------------------
\subsection{Induced extremes in the statistics of the coupled system} \label{sec:ext}

The study of extremes in dynamical systems has been the subject of extensive investigation (see \citep{lucarini_extremes_2016,ragone_computation_2018,galfi_large_2019} and references therein). We are interested here in the extremes of the slow coupled system for the following reasons. Intermittent episodes have been shown to reshape the geometry of the underlying attractor, altering its fractal dimension \cite{barone_structural_2025}.
Such structural changes leave a distinct signature in the long-term statistical properties of the system, and can be effectively probed through the lens of extreme value theory. Indeed, it is intuitive that a single chaotic burst embedded within an otherwise regular or weakly fluctuating signal can drastically alter the tail of the underlying probability distribution, motivating a systematic characterization of extremes across the parameter space explored in this study.

For both L63 and KSE, the extreme value analysis relies on an adaptive block length determined by the intrinsic decorrelation time \citep{von_storch_statistical_1999} $\tau_d$ of the response subsystem $\mathbf{X}_2$, rather than on a fixed block size \citep{coles_introduction_2001, lucarini_extremes_2016}. This choice accounts for the fact that the memory time of $\mathbf{X}_2$ varies substantially across the $(\tau,C)$ parameter space, so that a fixed block length would either violate the approximate independence of block maxima, when too short relative to $\tau_d$, or, when too long, drastically reduce the effective sample size available for the fit.

The decorrelation time $\tau_d$ is estimated from the autocorrelation function of a scalar observable of $\mathbf{X}_2$, computed after removing the sample mean and normalizing by the zero-lag value. It is defined as the first time at which the autocorrelation falls below the threshold $1/e$ \citep{von_storch_statistical_1999} and remains below this value for a minimum persistence window, a condition that prevents spurious detections due to short-lived oscillations around the threshold. Once $\tau_d$ is estimated, a long trajectory is generated for each parameter pair $(\tau,C)$ and, after discarding an initial transient, it is partitioned into blocks of length $\Delta T = a\,\tau_d$, with $a=2$, subject to a minimum block length of $20$~TU to guarantee a well-defined maximum even when $\tau_d$ is very small. The block maxima of the observable are then fitted to a Generalized Extreme Value (GEV) distribution by maximum likelihood estimation as implemented in the \texttt{genextreme} class of the \texttt{scipy.stats} module \citep{virtanen_scipy_2020}. Parameter combinations for which $\tau_d$ does not converge, {\it i.e.} the autocorrelation function never remains persistently below the $1/e$ threshold, or for which the block length does not yield a sufficient number of approximately independent block maxima, are excluded from the analysis and rendered as gray regions in the parameter maps. Altogether, these conditions lead to the exclusion of portions of the parameter space associated with laminar regimes, where the GEV estimation would be meaningless. For details on the numerical experiments see Table~\ref{tab:experiments}.

\paragraph{{\bf L63} -} The observable used to estimate $\tau_d$ and to extract the block maxima is the component $x_2(t)$ of the response subsystem. The same type of trajectory is used for the estimation of $\tau_d$ and for the extraction of the block maxima, and all parameter combinations are integrated over the full length. Combinations are excluded only a posteriori, when the resulting block length does not yield a sufficient number of approximately independent block maxima.

\paragraph{{\bf KSE} -} The observable used to estimate $\tau_d$ and to extract the block maxima is the global energy of the response field, $\varepsilon_2(t)$, used as a scalar proxy for $\mathbf{X}_2$. Owing to the high computational cost of the KSE, $\tau_d$ is estimated from a shorter, dedicated trajectory, independent from the longer one subsequently used for the extraction of the block maxima. In addition, an upper bound is enforced on $\Delta T$ prior to generating the long trajectory: when $\Delta T$ exceeds this bound, the parameter combination is directly excluded and the computationally expensive long trajectory is not generated at all.

\begin{figure*}[t]
    \centering
    \includegraphics[width=\textwidth]{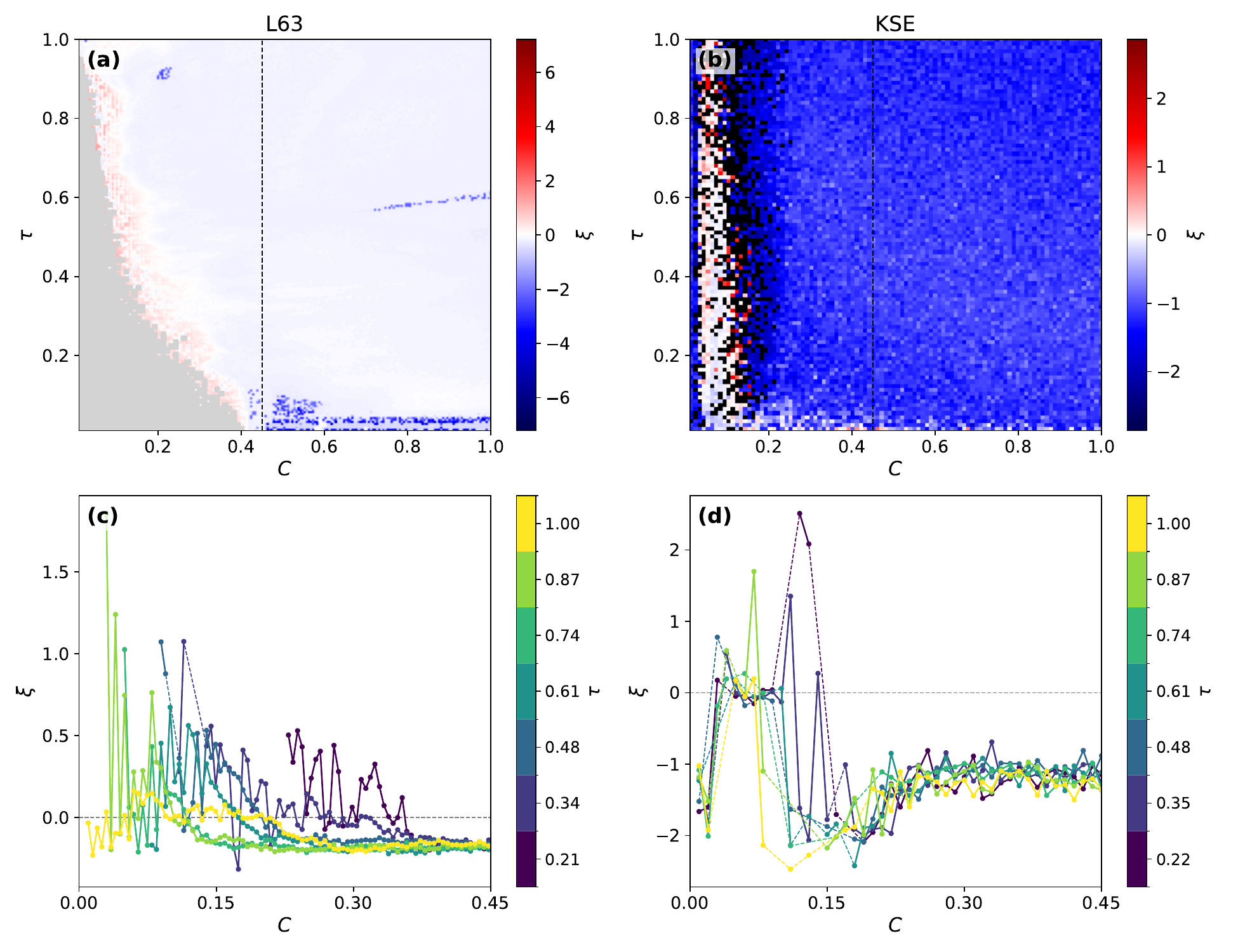}
    \caption{Shape parameter, $\xi$, of the GEV distribution fitted to the block maxima of the response subsystem. Top: $\xi$ over the $(\tau, C)$ parameter plane for (a) L63 and (b) KSE. Positive $\xi$ indicates a Fr\'echet-type, heavy-tailed distribution; $\xi\approx 0$ indicates a Gumbel-type distribution; negative $\xi$ indicates a Weibull-type, bounded distribution. Parameter combinations excluded from the analysis, either because the decorrelation time $\tau_d$ of the response could not be reliably estimated, or because the resulting block length did not yield a sufficient number of approximately independent block maxima, are shown in gray in (a) and in black in (b). The vertical dashed line marks $C=0.45$, the upper limit of the range shown in the bottom panels. Bottom: $\xi$ as a function of $C$, for $0\le C\le 0.45$ and five values of $\tau$ (color), for (c) L63 and (d) KSE. Gaps in the curves correspond to excluded parameter combinations and are bridged by thin dashed segments.}
    \label{fig:C2tau2}
\end{figure*}

\paragraph{{\bf Results} -} Figure~\ref{fig:C2tau2}a shows the GEV shape parameter $\xi$ estimated from the block maxima of $\mathbf{X}_2$ over the $(\tau, C)$ parameter space for the L63 system. A large portion of the parameter space falls within the gray region, where the autocorrelation of $\mathbf{X}_2$ remains too persistent for a well-defined $\tau_d$ to be extracted, and the convergence toward an appropriate GEV distribution is too slow \cite{Balakrishnan1995}. This behavior corresponds to regimes in which the response subsystem retains a strong coherence with its pre-existing limit cycle, resulting in trajectories that are either strictly periodic or exhibit only marginal deviations from it. Notably, this gray region extends broadly along the $\tau$ axis, up to values close to $\tau = 1$, while remaining confined to $C < 0.4$ along the coupling direction. This indicates that the loss of periodicity is primarily governed by the coupling strength rather than by the time-scale ratio, a mechanism already anticipated by the analysis of the variance of the maxima in Fig.~\ref{fig:C2tau1}. Immediately beyond this region, a transition zone emerges, characterized by $\xi>0$, indicative of a Fr\'echet-type, heavy-tailed distribution. This band is directly associated with the transmission of chaotic bursts to the forced limit cycle, which in turn reshapes the geometry of the attractor and, consequently, the statistics of its extremes. 

Most of the remaining parameter space, by contrast, is characterized by $\xi\approx0$, corresponding to a Gumbel-type distribution with exponentially decaying tails: this is the signature of the generic, fully developed chaotic regime of the forced L63, in which extremes are neither systematically enhanced by residual periodic coherence nor bounded by any saturation mechanism, and the statistics of $\mathbf{X}_2$ closely resemble that of the uncoupled, intrinsically chaotic attractor. 

Finally, the narrower regions of $\xi<0$, associated with a Weibull-type distribution with a finite upper endpoint, mark a further change of dynamical regime: here the coupling drives the response subsystem into a saturated state in which the amplitude of the extremes is effectively bounded. We hypothesize that this band is, at least in part, associated with the onset of phase synchronization between the two subsystems: a strong coupling constrains the phase of $\mathbf{X}_2$ to follow that of the driver, thereby restricting the accessible region of phase space and compressing the tail of the extreme value distribution relative to the weakly coupled, fully chaotic regime. We return to this point in Sect.~\ref{sec:phase}.

The KSE system (Fig.~\ref{fig:C2tau2}b) exhibits a markedly different behavior, consistent with what was already observed in the analysis of the variance of the maxima (Fig.~\ref{fig:C2tau1}b). A first region of negative $\xi$, indicative of a Weibull-type, bounded extreme value distribution, is followed by a transition into positive values, which becomes increasingly pronounced at larger $C$. This is followed by a narrow band of near-zero shape parameter, corresponding to a Gumbel-type distribution, whose width progressively shrinks as $\tau$ increases. Beyond this band, $\xi$ returns to positive values before finally settling into a broad region of markedly negative shape parameter, which occupies the largest portion of the remaining $(\tau, C)$ space. The positive-$\xi$ bands indicate regions where chaotic bursts originated in the driver are transmitted to the response field, reshaping the geometry of its attractor and enhancing the likelihood of extreme energy fluctuations. The narrow Gumbel-like band identifies a comparatively narrow window of fully developed chaotic dynamics in which extremes follow the generic exponential-tail behavior, without any residual influence of the underlying periodic or synchronized structure.

To highlight the behavior close to the transition, Fig.~\ref{fig:C2tau2}c,d show the shape parameter as a function of the coupling strength, $\xi(C)$, for five values of $\tau$ in the restricted range $0\le C\le 0.45$. The two systems display a qualitatively similar behavior. For small $C$, wherever $\xi$ is defined ({\it i.e.} outside the excluded gray/black region), the shape parameter takes negative values; as $C$ increases, it rises into a spike of positive values, and then returns to negative values at larger coupling. The positive band, corresponding to a Fr\'echet-type, heavy-tailed distribution, is associated with intermittent phenomena: in this range of $C$, the chaotic bursts transmitted from the driver dominate the tail of the distribution of the block maxima of $\mathbf{X}_2$. This behavior is consistent with the homogenisation results of \cite{gottwald_model_2021}, which show that intermittent fast dynamics act on the slow variables as a non-Gaussian, $\alpha$-stable (L\'evy) forcing, whose heavy tails are inherited by the slow dynamics. Although our systems are not in the asymptotic limit of infinite time-scale separation in which this result is derived, the Fr\'echet-type band provides a finite-$\tau$ counterpart of this mechanism. In our setting, however, heavy-tailed statistics appear only within a limited range of $C$ and give way to bounded statistics at stronger coupling. We interpret this as an indication that the L\'evy-like signature of the forcing dominates only when the coupling is strong enough to break the periodic coherence of the response, but not so strong that the response is constrained to follow the driver. In L63, the position of the positive spike shifts toward larger $C$ as $\tau$ decreases, following the boundary of the excluded region in Fig.~\ref{fig:C2tau2}a, whereas in KSE it remains confined to $C\lesssim 0.15$ for all the values of $\tau$ considered. The two systems differ instead in the level reached at larger coupling, with $\xi$ settling to small negative values in L63 and to markedly negative values, $\xi\approx -1$, in KSE.

Overall, these results support the use of the GEV shape parameter as an efficacious metric to diagnose regime transitions induced by unidirectional coupling, complementing and reinforcing the picture already drawn by the variance of the maxima. In both systems, the dependence of $\xi$ on the coupling strength $C$ is more pronounced than that on $\tau$, confirming $C$ as a primary control parameter governing the long-term statistics of the forced subsystem, and the transition into the intermittency-dominated regime is marked by a band of positive shape parameter, reflecting the heavy-tailed statistics induced by chaotic bursts propagating from the driver. Regions of negative shape parameter identify instead states in which the amplitude of the extremes becomes effectively bounded. The qualitatively different balance between these regimes in the two systems, with Gumbel-type statistics prevailing in L63 and the Weibull-type, bounded regime dominating the parameter space at strong coupling in KSE, points to system-specific differences in how coupling reorganizes the attractor geometry of the response subsystem, a point that is also explored through the lens of phase synchronization in Sect.~\ref{sec:phase}.

%% ====================================================================
\section{The role of the time-scale difference} \label{SEC:RESP} 

Delayed responses of slow components are typically encountered in slow--fast dynamical systems, a notable example being the coupling between ocean and atmosphere \cite{Deser2010}. In the stochastic climate framework introduced in \cite{Hasselmann1976,Frankignoul1977}, the slowly evolving ocean integrates the rapidly fluctuating atmospheric forcing. This separation of time scales can also give rise to genuine delayed feedbacks. A prominent example is the El Ni\~no--La Ni\~na dynamics, for which the propagation and reflection of equatorial Kelvin and Rossby waves provide a delayed oceanic feedback to the rapidly adjusting atmosphere \cite{Suarez1988,Tziperman1994}. Coupling-induced phase locking has also been found in El Ni\~no dynamics and in midlatitude ocean--atmosphere systems \cite{Gallego2001,Zaliapin2010,Feliks2011}. The characteristics, origins and impact of delayed responses are explored here in detail with the aid of our idealized slow-fast systems.

\begin{figure*}[h!]
    \centering
    \includegraphics[width=\linewidth]{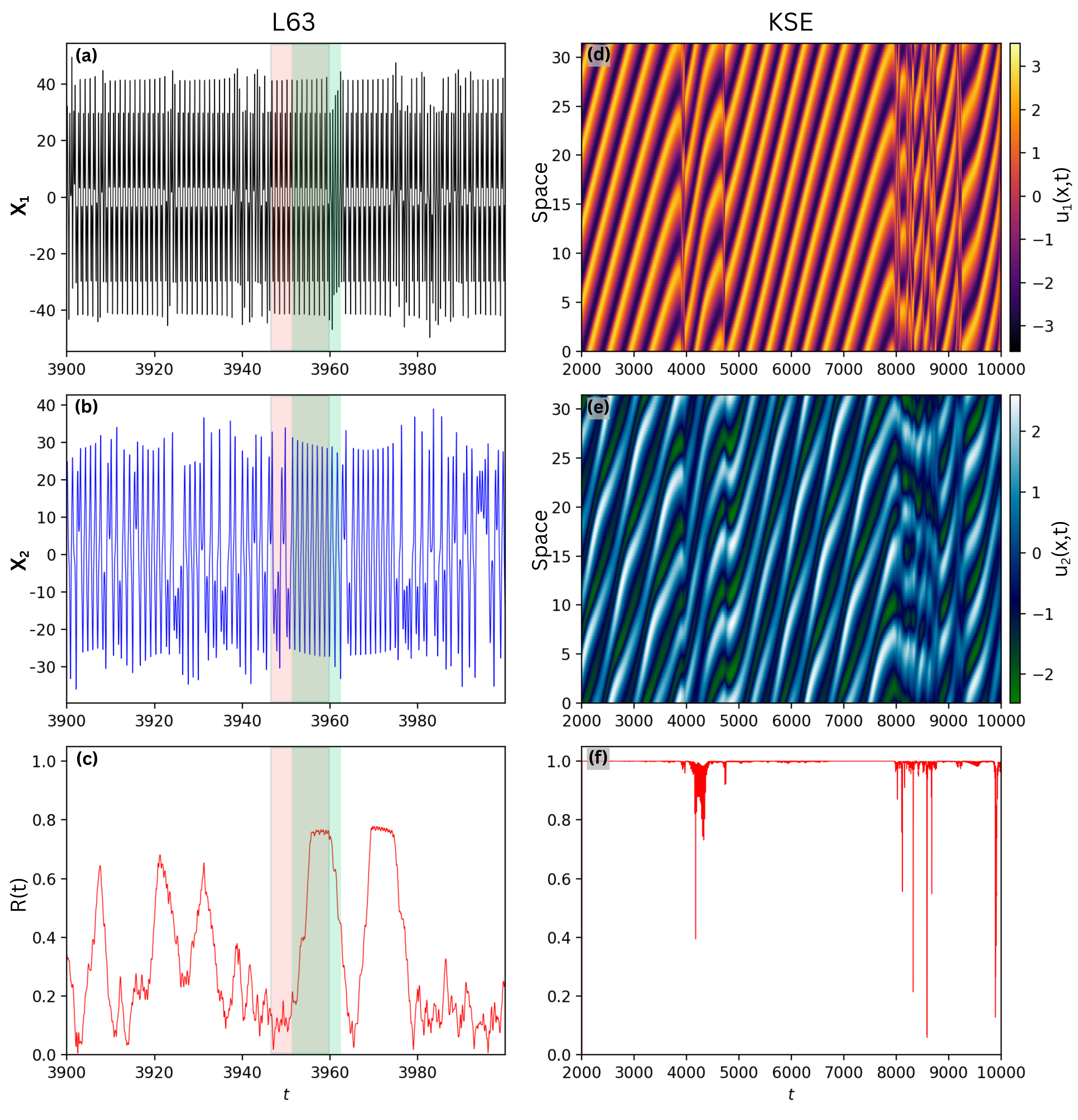}
    \caption{Local response delay and phase locking in the L63 (left column) and KSE (right column) systems. (a)--(c) L63 system: time series of the $x$-component of the driver $\mathbf{X}_1$ (a) and of the response $\mathbf{X}_2$ (b), with the driver's laminar onset (red shading) and the corresponding, delayed laminar onset in the response (green shading); (c) local, sliding-window phase-locking index $R(t)$. (d)--(f) KSE system: Hovm\"oller diagrams of the driver field $u_1(x,t)$ (d) and of the response field $u_2(x,t)$ (e); (f) local phase-locking index $R(t)$ computed from the Hilbert phases of the driver and response energies $\varepsilon_1(t)$ and $\varepsilon_2(t)$. Note the different time windows of the two columns, reflecting the much longer intermittent time scales of the KSE. Parameters are given in Table~\ref{tab:experiments}.}
    \label{fig:TS}
\end{figure*}

We investigate how the time-scale separation $\tau$ between the driver and the response affects the delay with which the slow subsystem responds to the intermittent bursts of the driver. Two diagnostics are used. The first is a local, time-resolved measure of phase coherence between the two subsystems. The instantaneous phase of each subsystem is extracted from the analytic signal associated with the (mean-removed) scalar observable $s_i(t)$, $i=1,2$,
\[
z_i(t) = s_i(t) + \mathrm{i}\,\tilde s_i(t), \qquad \phi_i(t) = \mathrm{unwrap}\big[\arg z_i(t)\big],
\]
with $\tilde s_i(t)$ the Hilbert transform \cite{rosenblum_phase_1996, pikovsky_synchronization_2001} of $s_i(t)$, computed numerically via the \texttt{hilbert} function of the \texttt{scipy.signal} module \citep{virtanen_scipy_2020}, with phase unwrapping performed using \\ \texttt{numpy.unwrap} \citep{harris_array_2020}. Rather than averaging the phase difference $\Delta\phi(t) = \phi_1(t) - \phi_2(t)$ over the whole trajectory, the local phase-locking index \cite{pikovsky_synchronization_2001} is computed over a sliding window of width $w$,
\[
R(t) = \Big| \big\langle e^{\mathrm{i}\Delta\phi(t')} \big\rangle_{t' \in [t-w,\,t]} \Big|.
\]
This provides a time-resolved measure of phase coherence, capturing transitions between synchronized and desynchronized episodes.

\begin{figure*}[t]
    \centering
    \includegraphics[width=0.9\linewidth]{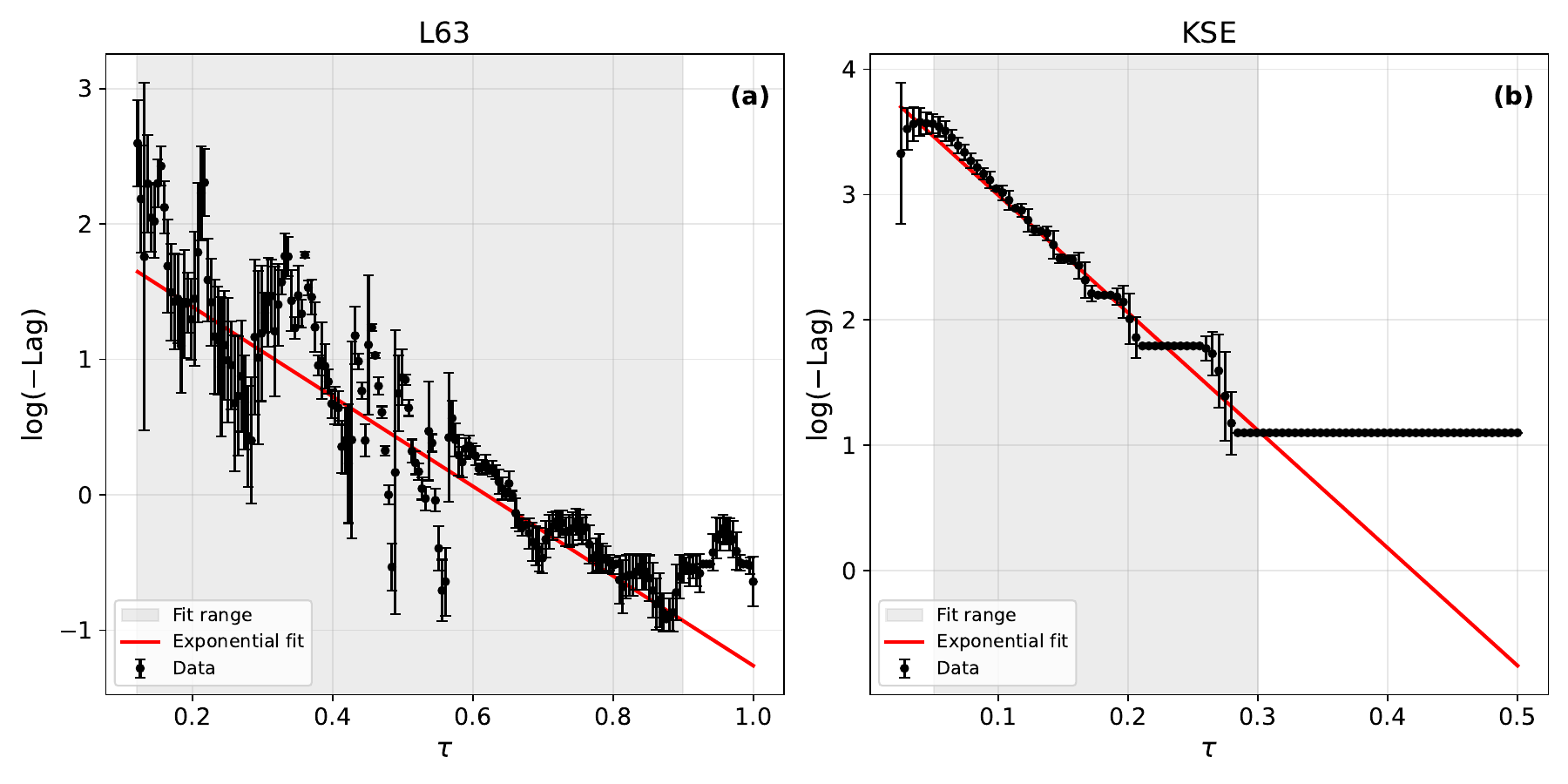}
    \caption{Mutual-information-based estimate of the optimal lag $l^*$ between driver and response, shown as $\log(-l^*)$ versus the time-scale separation $\tau$, for (a) L63 and (b) KSE. Error bars denote the standard deviation across an ensemble of independent realizations. The red line shows a weighted exponential fit performed over the shaded region; beyond $\tau\approx0.9$ (L63) and $\tau\approx0.3$ (KSE) the lag saturates and departs from the fitted exponential law.}
    \label{fig:scaling}
\end{figure*}

The second diagnostic quantifies the delay between driver and response through the time-lagged mutual information (MI),
\[
I(l) = I\big(s_1(t+l),\, s_2(t)\big), \qquad l \in [-l_{\max}, l_{\max}],
\]
estimated from the joint histogram of the two (min--max normalized) signals, rather than with the $k$-nearest-neighbour estimator of \cite{kraskov_estimating_2004}. For each $\tau$, at fixed coupling strength $C$, the optimal lag $l^{*}(\tau)$ is the value of $l$ maximizing $I(l)$ (after light Gaussian smoothing of the MI curve). A negative $l^{*}$ indicates that the response at $t$ best matches the driver at the earlier time $t+l^{*}$, {\it i.e.}, that the driven system responds with a delay of $|l^{*}|$. To estimate both the mean lag and its uncertainty, the procedure is repeated over an ensemble of independent realizations obtained by slightly perturbing the initial condition: the optimal lag at each $\tau$ is the ensemble mean $\langle l^{*}(\tau) \rangle$, with the ensemble standard deviation as the associated error. For details on the numerical experiments see Table~\ref{tab:experiments}.

\paragraph{{\bf L63} -} The observable is the $x$-component of each subsystem, $s_i(t) = x_i(t)$.

\paragraph{{\bf KSE} -} Since the coupling acts on the extended field rather than on a single scalar variable, the phase is extracted not from the raw field but from its energy, defined as in the previous sections as the instantaneous $L^2$ norm of the driver and response fields, $s_i(t) = \varepsilon_i(t) = \lVert u_i(x,t)\rVert_{L^2}$. The Hilbert phase, $R(t)$ and the MI are then computed following the same procedure.

\paragraph{{\bf Results} -} Figure~\ref{fig:TS}a,b shows the $x$-components of the L63 driver $\mathbf{X}_1$ and response $\mathbf{X}_2$. When a laminar phase is initiated in the driver (red shaded), the onset of the corresponding laminar phase in the response (green shaded) is delayed by a finite lag $\delta t$. The oscillations of the response then persist for some time after the driver has left its laminar phase and re-entered a chaotic burst: the slow subsystem retains a memory of the previous forcing episode beyond the duration of the driving laminar phase itself. This delayed response is quantified by $R(t)$ in Fig.~\ref{fig:TS}c: $R(t)$ grows progressively once the driver has left the chaotic burst and, after a transient, the two oscillators lock into a synchronized regime in which the phase difference remains approximately constant and $R(t)$ saturates around $0.8$, before dropping sharply as soon as the driver re-enters a new chaotic burst.

For the KSE, Fig.~\ref{fig:TS}d,e shows the Hovm\"oller diagrams of the driver and response fields, and Fig.~\ref{fig:TS}f the corresponding $R(t)$. The local phenomenology is markedly different: $R(t)$ exhibits sharp, spike-like drops confined to the chaotic bursts of the driver, while it remains close to saturation throughout each laminar episode.

We now turn to the dependence of the delay on $\tau$ (Fig.~\ref{fig:scaling}). For L63 (Fig.~\ref{fig:scaling}a), the optimal lag decreases exponentially as $\tau$ approaches unity, {\it i.e.}, as the time scales of the driver and the response become comparable. A weighted linear fit of $\log(-l^{*})$ versus $\tau$ is performed over the shaded region, in which the scaling is well described by a single exponential law. The estimate is considerably less smooth than its KSE counterpart, with substantial scatter and large error bars across the whole range of $\tau$; the lag does not saturate until $\tau \approx 0.9$, where it appears to increase slightly again before the equal-time-scale limit is reached.

% CHECK: verify that the plateaus are not an artefact of the lag resolution (output time step); if they are, drop the devil's staircase sentence
For the KSE (Fig.~\ref{fig:scaling}b), the same exponential-type decay is observed, but the estimate is markedly smoother, with much smaller ensemble errors. In several instances the optimal lag coincides exactly across neighboring values of $\tau$ (the multiple plateaus in the figure), and the lag saturates already before $\tau \approx 0.3$, staying constant afterward. These plateaus, over which the lag remains locked across finite intervals of $\tau$, are reminiscent of the mode-locking steps of a devil's staircase \citep{jensen_complete_1983, pikovsky_synchronization_2001}.

Taken together, these results support the interpretation that the exponential scaling of the response delay with $\tau$ is a robust feature shared by both coupled type-I intermittent systems considered here, rather than a peculiarity of either model. In both the low-dimensional (L63) and the spatially extended (KSE) case, the response of the slow subsystem to the driver's intermittent bursts follows a characteristic exponential law, while the specific form of the deviations from this law (residual variability in L63, sharp plateaus in KSE) reflects the different nature of the two systems.

%% ====================================================================
\section{Phase locking} \label{sec:phase}

We now turn to the phase-locking index \cite{pikovsky_synchronization_2001,rosenblum_phase_1996} averaged over the whole trajectory,
\[
\overline{R} = \left| \left\langle e^{\mathrm{i} \Delta \phi(t)} \right\rangle \right|,
\]
with the phases computed as in Sect.~\ref{SEC:RESP}. $\overline{R} \in [0,1]$ measures the overall degree of phase coherence between the driver and the response subsystem: larger $\overline{R}$ implies stronger phase coherence. For details on the numerical experiments see Table~\ref{tab:experiments}.

\paragraph{{\bf L63} -} $\overline{R}$ is computed over the $(\tau,C)$ plane with the intermittent driver. As a control experiment, the same scan is repeated with the intermittent driver replaced by a purely periodic one.

\paragraph{{\bf KSE} -} Given the higher computational cost of a full two-dimensional scan of the $(\tau,C)$ plane, $\overline{R}$ is evaluated only along three directions of the parameter space, obtained by fixing an angle $\theta$ and varying the radial distance from the origin, so that $\tau = C/\tan\theta$ along each direction.

\begin{figure}[h!]
    \centering
    \includegraphics[width=0.9\columnwidth]{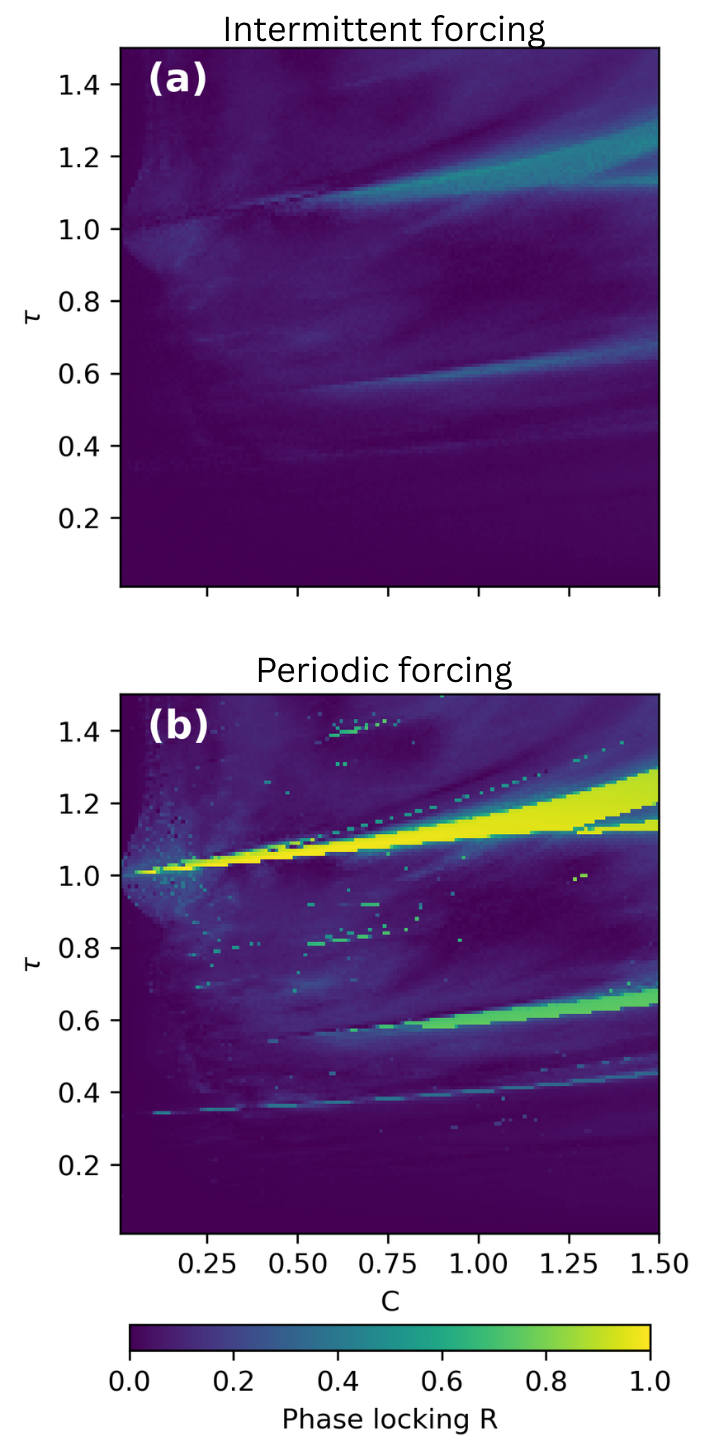}
    \caption{Time-averaged phase-locking index $\overline{R}$ over the $(\tau,C)$ parameter plane for the coupled L63 system, under (a) intermittent or (b) purely periodic forcing.}
    \label{fig:C2tau3}
\end{figure}

\begin{figure}[h!]
    \centering
    \includegraphics[width=\columnwidth]{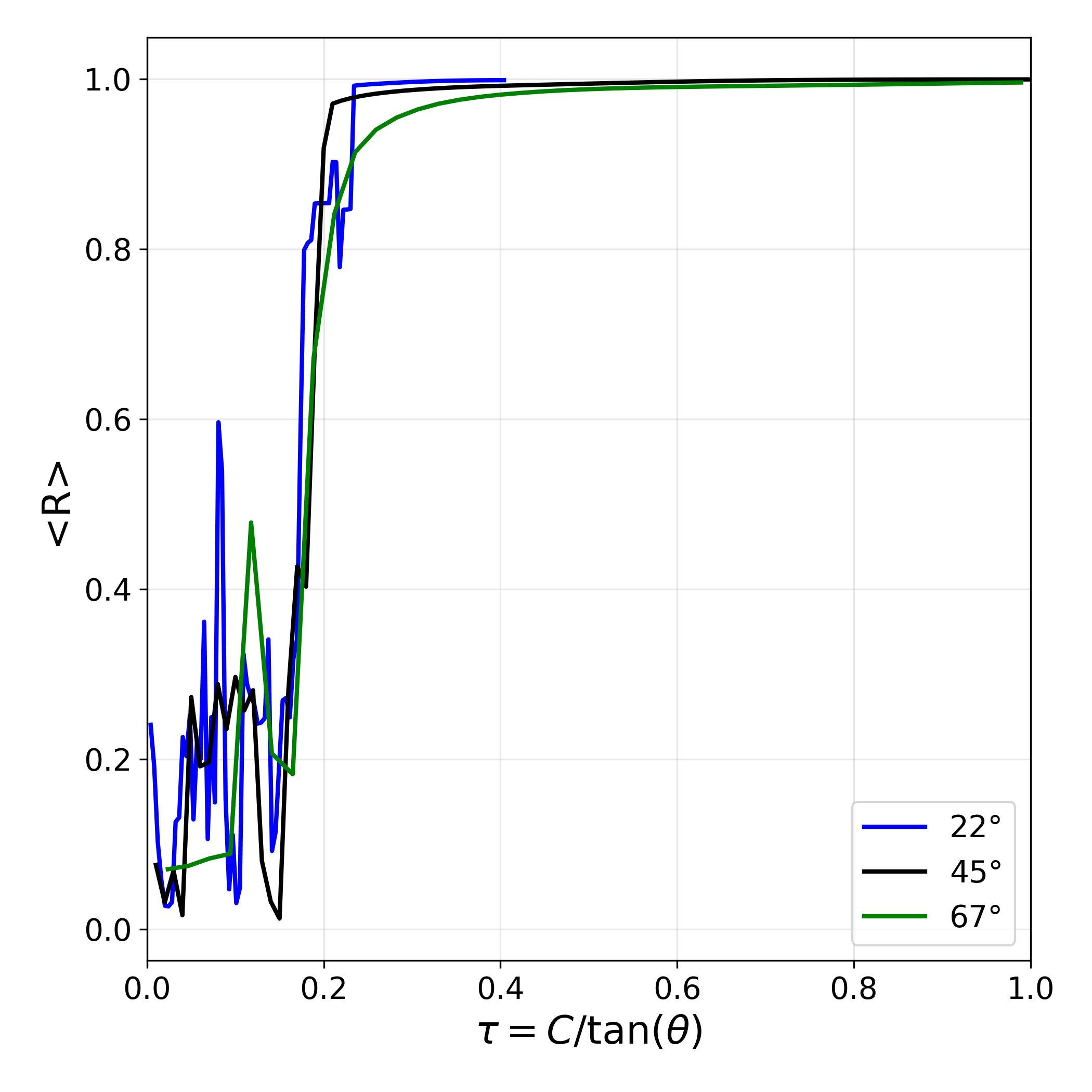}
    \caption{Time-averaged phase-locking index $\overline{R}$ for the KSE system, evaluated along three directions of the $(\tau,C)$ plane parametrized by $\tau = C/\tan\theta$, for $\theta = 22^\circ$ (blue), $45^\circ$ (black), and $67^\circ$ (green). All three curves saturate shortly before $\tau\approx0.3$ and collapse onto a common curve representative of the driver system.}
    \label{fig:phse_KSE}
\end{figure}

\paragraph{{\bf Results} -} Figure~\ref{fig:C2tau3}a shows $\overline{R}$ as a function of $C$ and $\tau$ for the L63 system under intermittent forcing. Faint bands with higher $\overline{R}$ than elsewhere are visible, marking regions where the driver and the driven systems are in phase. As already observed in the local, time-resolved analysis ({\it cf.} Fig.~\ref{fig:TS}), we attribute this behavior to synchronization triggered by the periodic (laminar) segments of the intermittent signal. This is scrutinized in Fig.~\ref{fig:C2tau3}b, where the intermittent driver is replaced by a periodic forcing: the same regions of phase coherence appear, but with a much higher index, $\overline{R} \approx 1$. This reinforces the picture outlined above: under intermittent forcing, phase locking is only partial and disrupted by the chaotic bursts, whereas under purely periodic forcing the two attractors remain phase-locked at all times. Moreover, the coincidence of the phase-coherent regions for the same values of $C$ and $\tau$ suggests that they originate from the same synchronization mechanism. Synchronization induces a structural change in the response attractor, from a period-2 oscillation, typical of the uncoupled L63 in this range of $\rho$, to an attractor exhibiting a single dominant period. We refer the reader to the Supplementary Material for a more extensive discussion.

For the KSE, Fig.~\ref{fig:phse_KSE} shows the curves of $\overline{R}$ as a function of $\tau = C/\tan\theta$ for the three directions. After an initial region of low, noisy values, all three curves saturate shortly before $\tau \approx 0.3$, consistent with the saturation threshold identified in the response-delay scaling of Sect.~\ref{SEC:RESP}, and they collapse onto a common curve. This indicates that, in the KSE and within the range explored, $\overline{R}$ is controlled by $\tau$ alone, its dependence on $C$ along each direction being negligible. Moreover, the fact that saturation is reached already at relatively small $\tau$, and for all three directions, suggests that the driver's dynamics is imprinted onto the response subsystem considerably more efficiently in the KSE than in the L63, where the transition to phase locking is smoother and confined to narrower bands in parameter space.

%% ====================================================================
\section{Conclusion} \label{sec:concl}

\subsection{Summary}
This work investigated how the intermittent behavior of a fast subsystem $\mathbf{X}_1$ is transmitted to, and reorganizes the dynamics of, a slower subsystem $\mathbf{X}_2$ that, in the absence of forcing, would settle onto a stationary, periodic regime. The two subsystems form a skew-product system: the coupling is strictly unidirectional, so that $\mathbf{X}_1$ evolves autonomously and drives $\mathbf{X}_2$ without any feedback from the response, and the separation between their characteristic time scales is controlled by the parameter $\tau$. We studied this setting in two structurally different models of increasing complexity, based respectively on the Lorenz-63 system and on the spatially extended Kuramoto--Sivashinsky equation.

Several diagnostics were used to assess the strength, type and shape of the impact of the fast intermittent subsystem on the slow one. These include the Wasserstein distance between the induced and reference attractors, the variance of the response's local maxima, the power spectral density of the response relative to the driver, the shape parameter of the fitted GEV distribution, and the phase-locking index $\overline{R}$. All the diagnostics consistently picture the same mechanism: as the coupling strength $C$ and/or the time-scale parameter $\tau$ are increased, the response subsystem is progressively pulled away from its own natural attractor, and its statistical, spectral and extreme-value properties are reshaped toward those of the driver. Crucially, this transfer of information is not unbounded: in both systems, the induced perturbation of $\mathbf{X}_2$ saturates once the driver's burst frequency exceeds a certain threshold, indicating that the response subsystem possesses an intrinsic capacity to absorb intermittent forcing, beyond which no additional distortion of its attractor is produced.

A central result is the identification of phase synchronization as the mechanism responsible for several of the regime transitions observed in the L63 case. The stripe-like structures found in the variance of the maxima and in the time-averaged phase-locking maps $\overline{R}(\tau,C)$ occupy the same regions of the $(\tau,C)$ plane, and the bounded, Weibull-type regime of the GEV shape parameter is, at least in part, associated with the same synchronized states. To identify the origin of these bands, we performed a control experiment in which the intermittent driver is replaced by a purely periodic one. In this configuration the phase of the driver is well defined at all times and is no longer interrupted by turbulent bursts. If the bands reflected phase locking between the two subsystems, they should persist and become complete in the absence of bursts; if they arose from other features of the intermittent signal, they should disappear. We observe the former: the same bands saturate at $\overline{R} \approx 1$. This indicates that the bands arise from phase locking, and that the values $\overline{R} < 1$ found with the intermittent driver reflect phase slips induced by the bursts, which repeatedly break an otherwise stable locking. This synchronization carries a direct structural signature, driving the response attractor from a period-2 oscillation to a single-period orbit, and it also modulates the local, time-resolved dynamics: the local index $R(t)$ grows during the driver's laminar phases and collapses during its chaotic bursts.

Using a mutual-information-based estimate of the optimal lag, we show that the response delay follows a common exponential scaling law with $\tau$ in both systems, decreasing as the two time scales become comparable. The robustness of this scaling across two structurally different models suggests that it is a manifestation of a general property of how information propagates across time-scale-separated, intermittently forced systems.

At the same time, the comparison between L63 and KSE shows that the way this general mechanism works is also shaped by the internal structure of each system. In L63, the coupling strength $C$ emerges as the dominant control parameter for the amplitude-based diagnostics, governing the transition to chaos, the loss of periodicity, and the statistics of extremes, while $\tau$ mainly controls the reshaping of the response spectrum. Extremes in the fully forced regime are generically of Gumbel type, and the delay-scaling curve displays considerable scatter, saturating only close to $\tau \approx 0.9$. The KSE system instead exhibits a markedly more efficient and less noisy synchronization: it saturates already before $\tau \approx 0.3$, both in its delay scaling and in its phase-locking curves, which further collapse onto a single curve when plotted against $\tau$ along the three directions $\tau = C/\tan\theta$, indicating that phase locking is controlled by the time-scale separation rather than by the coupling strength. KSE extremes, moreover, are generically bounded (Weibull-type) rather than Gumbel-type at strong coupling, suggesting that saturation, rather than fully developed chaos, is the typical outcome of strong forcing in this higher-dimensional, spatially extended setting.

Taken together, these results suggest a coherent answer to the question posed at the outset: intermittent information originating in a fast subsystem is transmitted to a slower one primarily by reshaping its attractor and its extreme-event statistics up to a saturation point set by the coupling strength and the time-scale separation. Phase synchronization is the specific mechanism through which this transmission organizes itself into distinct dynamical regimes, imprinting a structural signature (attractor topology, delay, extreme-value statistics) on the response subsystem. Beyond their relevance for multiscale nonlinear dynamics in general, these findings lend some quantitative support to the conceptual climate motivation outlined in the Introduction: an intermittent, atmospheric-like component forcing a near-stationary, oceanic-like mode may be expected to destabilize it, alter its extreme-event statistics, and induce a lagged, synchronization-mediated response, with the strength and delay of this response governed in a systematic, and to some extent universal, way by the coupling strength and the separation of time scales between the two components.

\subsection{Future perspectives}

The present study is limited to the unidirectional, feedback-free configuration, chosen deliberately to isolate the forcing mechanism in its simplest form. A natural extension is to reintroduce feedback from the slow to the fast subsystem, and to assess whether the synchronization-driven regimes identified here persist, are enhanced, or are suppressed once the response is allowed to act back on its driver. Likewise, a natural direction for future work is to test whether the mechanisms identified here, namely the exponential delay scaling and the synchronization-driven regime transitions, extend to intermittent phenomena in systems of increasing complexity. Identifying analogous phase-locking signatures in higher-dimensional, more realistic multiscale models would help establish whether the picture emerging here reflects a genuinely general property of intermittently forced fast--slow systems, rather than a feature specific to the two idealized models considered. Such an extension would also provide a more direct test of the climate-oriented motivation of this work, clarifying whether the mechanisms identified here can shed light on the role that intermittent signals, such as fast atmospheric variability, play in shaping the statistics and predictability of slower components of the Earth system.

Another important way forward is to develop additional tools for detecting transitions between regimes and phase locking. A potential direction is to understand the transitions and phase locking by exploring the preferential paths between the different regimes and the role played by unstable periodic orbits, see \cite[e.g.][]{Hamilton2025}, or using characteristic Lyapunov vectors as in \cite{barone_structural_2025}. This will potentially allow the construction of robust precursors of transitions.

%% ====================================================================
\appendix

\section{Sliced Wasserstein Distance}
\label{app:sliced-wd}

The sliced Wasserstein distance (SWD) reduces a $d$-dimensional optimal transport problem to a family of one-dimensional ones \cite{bonneel2015sliced}. Given two probability measures $\mu$ and $\nu$ on $\mathbb{R}^d$ and a unit direction $\theta\in\mathbb{S}^{d-1}$, let $\theta_{\#}\mu$ denote the distribution of the scalar projection of samples from $\mu$ onto $\theta$. The SWD is the average, over all directions on the unit sphere and with respect to the uniform measure $\sigma$ on $\mathbb{S}^{d-1}$, of the one-dimensional Wasserstein distance $W$ of order one between these projected marginals:
\begin{equation}
  \mathrm{SW}(\mu,\nu) = \int_{\mathbb{S}^{d-1}} W\big(\theta_{\#}\mu,\, \theta_{\#}\nu\big)\, d\sigma(\theta).
\end{equation}
In one dimension, $W$ admits a closed form in terms of quantile functions, and for empirical samples it reduces to sorting \cite{santambrogio2015optimal}. The SWD thus replaces one difficult, high-dimensional transport problem with an average over many easy, one-dimensional ones.

In practice, the integral over $\mathbb{S}^{d-1}$ is approximated by drawing $n_{\mathrm{proj}}$ directions $\theta_1,\dots,\theta_{n_{\mathrm{proj}}}$ uniformly on the sphere (obtained by normalizing standard Gaussian vectors in $\mathbb{R}^d$) and averaging:
\begin{equation}
  \mathrm{SW}(\mu,\nu) \approx \frac{1}{n_{\mathrm{proj}}} \sum_{i=1}^{n_{\mathrm{proj}}} W\big(\theta_i\!\cdot\! X,\, \theta_i\!\cdot\! Y\big),
\end{equation}
where $X$ and $Y$ denote the empirical samples from $\mu$ and $\nu$, and $\theta_i\cdot X$ denotes the one-dimensional set of scalar projections of every sample in $X$ onto $\theta_i$.

In this work, this procedure is applied twice: to the standardized response-subsystem state vector $\mathbf{X}_2=(x_2,y_2,z_2)$ in the Lorenz-63 case ($d=3$), and to point clouds in the $(u_x,u_{xx})$ phase plane in the Kuramoto--Sivashinsky case ($d=2$). In both cases $n_{\mathrm{proj}}=100$ random directions are used; each of the resulting one-dimensional Wasserstein distances is computed exactly in closed form, and the $n_{\mathrm{proj}}$ values are averaged to obtain the distance $\mathcal{D}$ between each swept distribution and its corresponding reference distribution. 

%% ====================================================================
\section*{Author Declarations}
\subsection{Conflict of Interest}
The authors have no conflicts to disclose.

\subsection{Author Contributions}
\textbf{A. Barone}: Conceptualization (lead); Formal analysis (lead); Investigation (lead); Methodology (lead); Software (lead); Validation (lead); Visualization (lead); Writing – original draft (lead); Writing – review \& editing (lead). \textbf{A. Carrassi}: Conceptualization (supporting); Formal analysis (supporting); Investigation (supporting); Methodology (supporting); Validation (supporting); Writing – original draft (supporting); Writing – review \& editing (supporting). \textbf{J. Demaeyer}: Conceptualization (supporting); Investigation (supporting); Methodology (supporting); Validation (supporting); Writing – review \& editing (supporting). \textbf{S. Vannitsem}: Conceptualization (supporting); Investigation (supporting); Methodology (supporting); Validation (supporting); Writing – review \& editing (supporting).

\section*{Data availability}
The data that support the findings of this study are available from the corresponding author upon reasonable request.

\section*{Acknowledgements}
A.B. and A.C. have been funded as members of the Scale-Aware Sea Ice Project (SASIP) supported by Grant No. G-24-66154 of Schmidt Sciences, LLC—a philanthropy that propels scientific knowledge and breakthroughs toward a thriving world.\textbf{}

\bibliographystyle{model1-num-names}
\bibliography{cas-refs}

\end{document}